\documentstyle[epsfig]{mn}

\newif\ifAMStwofonts

\def\t{\thinspace}

\def \rns {R_{\rm ns}}
\def \egret {{\it EGRET}}
\def \glast {{\it GLAST}}
\def \al {$\alpha$}
\def \ze {$\zeta$}

\def  \del {\Delta^{\rm peak}}
\def  \delcr {\Delta^{\rm peak}_{\rm cr}}

\def  \et  {\varepsilon_{\rm turn}}
\def  \ee  {\varepsilon_{\rm esc}}
\def  \esr {\varepsilon_{\rm sr}}
\def  \ecr {\varepsilon_{\rm cr}}
\def  \bcr {B_{\rm crit}}
\def  \gper {\gamma_\perp}

\def  \eb   {\varepsilon_{B}}
\def  \ect  {\varepsilon_{\rm ct}}
\def  \r90  {R_{90}}
\def  \epm  {e^\pm}
\def  \bpc  {B_{\rm pc}}
\def  \rpc  {r_{\rm pc}}
\def  \thg  {\vartheta_\gamma}

\def\GeV{~\rm{GeV}}
\def\MeV{~\rm{MeV}}

\def\G{~\rm{G}}

\newbox\grsign \setbox\grsign=\hbox{$>$} \newdimen\grdimen \grdimen=\ht\grsign
\newbox\simlessbox \newbox\simgreatbox \newbox\simpropbox
\setbox\simgreatbox=\hbox{\raise.5ex\hbox{$>$}\llap
     {\lower.5ex\hbox{$\sim$}}}\ht1=\grdimen\dp1=0pt
\setbox\simlessbox=\hbox{\raise.5ex\hbox{$<$}\llap
     {\lower.5ex\hbox{$\sim$}}}\ht2=\grdimen\dp2=0pt
\setbox\simpropbox=\hbox{\raise.5ex\hbox{$\propto$}\llap
     {\lower.5ex\hbox{$\sim$}}}\ht2=\grdimen\dp2=0pt
\def\simgreat{\mathrel{\copy\simgreatbox}}
\def\simless{\mathrel{\copy\simlessbox}}

\def  \above  #1#2{$\begin{array}{c} #1 \\
                                     \noalign{\vskip 5pt}
                                     #2 \end{array}$}

\ifoldfss

  \newcommand{\bld}[1] {{\bf #1}}
  \ifCUPmtlplainloaded \else
    \NewTextAlphabet{textbfit} {cmbxti10} {}
    \NewTextAlphabet{textbfss} {cmssbx10} {}
    \NewMathAlphabet{mathbfit} {cmbxti10} {} 
    \NewMathAlphabet{mathbfss} {cmssbx10} {} 
  \fi
  \ifAMStwofonts
    \ifCUPmtlplainloaded \else
      \NewSymbolFont{upmath} {eurm10}
      \NewSymbolFont{AMSa} {msam10}
      \NewMathSymbol{\upi}     {0}{upmath}{19}
      \NewMathSymbol{\umu}     {0}{upmath}{16}
      \NewMathSymbol{\upartial}{0}{upmath}{40}
      \NewMathSymbol{\leqslant}{3}{AMSa}{36}
      \NewMathSymbol{\geqslant}{3}{AMSa}{3E}

      \let\leq=\leqslant 
      \let\geq=\geqslant \let\ge=\geqslant
    \fi
  \fi
\fi 

\ifnfssone
  \newmathalphabet{\mathit}
  \addtoversion{normal}{\mathit}{cmr}{m}{it}
  \addtoversion{bold}{\mathit}{cmr}{bx}{it}

  \newcommand{\bld}[1] {\mathbf{#1}}
  \newmathalphabet{\mathbfit} 
  \addtoversion{normal}{\mathbfit}{cmr}{bx}{it}
  \addtoversion{bold}{\mathbfit}{cmr}{bx}{it}
  \newmathalphabet{\mathbfss} 
  \addtoversion{normal}{\mathbfss}{cmss}{bx}{n}
  \addtoversion{bold}{\mathbfss}{cmss}{bx}{n}
  \ifAMStwofonts
    \ifCUPmtlplainloaded \else
      \UseAMStwoboldmath
      \makeatletter
      \new@mathgroup\upmath@group
      \define@mathgroup\mv@normal\upmath@group{eur}{m}{n}
      \define@mathgroup\mv@bold\upmath@group{eur}{b}{n}
      \edef\UPM{\hexnumber\upmath@group}
      \new@mathgroup\amsa@group
      \define@mathgroup\mv@normal\amsa@group{msa}{m}{n}
      \define@mathgroup\mv@bold\amsa@group{msa}{m}{n}
      \edef\AMSa{\hexnumber\amsa@group}
      \makeatother
      \mathchardef\upi="0\UPM19
      \mathchardef\umu="0\UPM16
      \mathchardef\upartial="0\UPM40
      \mathchardef\leqslant="3\AMSa36
      \mathchardef\geqslant="3\AMSa3E

      \let\leq=\leqslant 
      \let\geq=\geqslant \let\ge=\geqslant
    \fi
  \fi
\fi 

\ifnfsstwo

  \newcommand{\bld}[1] {\mathbf{#1}}
  \DeclareMathAlphabet{\mathbfit}{OT1}{cmr}{bx}{it}
  \SetMathAlphabet\mathbfit{bold}{OT1}{cmr}{bx}{it}
  \DeclareMathAlphabet{\mathbfss}{OT1}{cmss}{bx}{n}
  \SetMathAlphabet\mathbfss{bold}{OT1}{cmss}{bx}{n}
  \ifAMStwofonts
    \ifCUPmtlplainloaded \else
      \DeclareSymbolFont{UPM}{U}{eur}{m}{n}
      \SetSymbolFont{UPM}{bold}{U}{eur}{b}{n}
      \DeclareSymbolFont{AMSa}{U}{msa}{m}{n}
      \DeclareMathSymbol{\upi}{0}{UPM}{"19}
      \DeclareMathSymbol{\umu}{0}{UPM}{"16}
      \DeclareMathSymbol{\upartial}{0}{UPM}{"40}
      \DeclareMathSymbol{\leqslant}{3}{AMSa}{"36}
      \DeclareMathSymbol{\geqslant}{3}{AMSa}{"3E}

      \let\leq=\leqslant 
      \let\geq=\geqslant \let\ge=\geqslant
    \fi
  \fi
\fi 

\ifCUPmtlplainloaded \else
  \ifAMStwofonts \else 
    \def\upi{\pi}
    \def\umu{\mu}
    \def\upartial{\partial}
  \fi
\fi

\title[Model of peak separation]
  {Model of peak separation in the gamma lightcurve of the Vela pulsar}
\author[J.~Dyks and B.~Rudak]
  {J.~Dyks\thanks{E-mail: jinx@ncac.torun.pl (JD); bronek@camk.edu.pl (BR)}
   and B.~Rudak\raise5pt\hbox{$\star$}\\
  Nicolaus Copernicus Astronomical Center,
Rabia{\'n}ska 8, 87-100 Toru{\'n}, Poland
}
\date{Accepted . Received }
\pagerange{\pageref{firstpage}--\pageref{lastpage}}
\pubyear{1994}

\def\LaTeX{L\kern-.36em\raise.3ex\hbox{a}\kern-.15em
    T\kern-.1667em\lower.7ex\hbox{E}\kern-.125emX}

\begin{document}

\label{firstpage}

\maketitle

\begin{abstract}
The separation $\del$ between two peaks in the gamma-ray pulse profile is 
calculated as a function of energy 
for several polar cap models with curvature-radiation-induced cascades. 
The Monte Carlo results are interpreted with the help of
analytical approximations and discussed in view of 
the recent data analysis for the Vela pulsar \cite{kanbach}.
We find that the  behaviour of $\del$
as a function of photon energy $\varepsilon$
depends primarily on local values of the magnetic field, $B_{\rm local}$, 
in the region where electromagnetic cascades develop.
For low values of $B_{\rm local}$ ($< 10^{12}$ G), $\del(\varepsilon)$ is kept constant.
However, for stronger magnetic fields ($\ga 10^{12}$ G) in the hollow-column model
$\del$ decreases with increasing photon energy
at a rate dependent on maximum energy of beam particles as well as on 
viewing geometry.
There exists a critical photon energy $\et$ above which 
the relation $\del(\varepsilon)$ changes drastically: for $\varepsilon > \et$,  
in hollow-column models the separation $\del$
increases (whereas in filled-column model it decreases) rapidly with increasing $\varepsilon$, at a rate  
of $\sim 0.28$ of the total phase per decade of photon
energy. The existence of critical energy $\et$ is a direct consequence of one-photon magnetic absorption effects.
In general, $\et$ is located close to 
the high-energy cutoff of the spectrum, thus photon statistics at $\et$ should be very low.
That will make difficult to verify an existence of $\et$  in real gamma-ray pulsars. 
Spectral properties of the Vela pulsar would favour those models which 
use low values of magnetic
field in the emission region ($B_{\rm local} \simless 10^{11}$ G) which in turn implies a constant value
of the predicted $\del$ within \egret\ range.
\end{abstract}


\section{Introduction}

Two prominent peaks are a characteristic feature 
of gamma-ray pulse shapes in the three brightest out of seven
gamma-ray pulsars detected so far: Crab (PSR B0531+21), Vela (PSR B0833-45), 
and Geminga (J0633+1746).
Phase separation between the two peaks is very large in
each case, 
in the range between 0.4 and 0.5 (e.g. Fierro, Michelson \& Nolan 1998).
The separation, which we denote by $\Delta^{\rm peak}$,
was determined with photons from the entire energy range of \egret.

Kanbach \shortcite{kanbach} suggested that the separation  $\Delta^{\rm peak}$ in Vela might be energy dependent. 
The effect would be of the order of a few percent or less. The plot 
of the phase separation against energy (fig.2 of Kanbach 1999) 
shows that $\Delta^{\rm peak}$ decreases by about $5\%$ 
over 20 energy intervals covering the range between $\sim 50\MeV$ and $\sim 9\GeV$. 
The scatter of points is, however, large enough for this result still to be
consistent with the separation staying at a constant level of $0.43$, especially 
when one rejects two energy intervals: of the lowest and the highest value. 

Such effects as suggested by Kanbach can be justified qualitatively, at least within polar cap scenarios.
Their origin may be different at different energy ranges, and their magnitude may vary as well.
For example, Miyazaki \& Takahara \shortcite{miyazaki} found dramatic changes in peak-to-peak phase separation
due to magnetic absorption effects in their attempts to model the Crab pulse shapes. Their numerical
calculations were performed with low photon energy resolution for a model with homogeneous polar
cap, and instant acceleration.

This new aspect of studying the HE properties of pulsars is potentially attractive.
The problem of poor photon statistics should become
less essential with future high-sensitivity missions like \glast.
Then any well established empirical relation between the phase separation $\Delta^{\rm peak}$ 
and the photon energy $\varepsilon$ (including $\Delta^{\rm peak} = const$)
may help to discriminate in favour of some particular models of pulsar activity.

In this context we present a model of the peak-to-peak phase separation 
in the gamma-ray lightcurve of Vela
and confront it with the results of Kanbach \shortcite{kanbach}.
Our aim is to present properties of $\Delta^{\rm peak}(\varepsilon)$
predicted by the polar cap model with curvature (CR) and synchrotron (SR) radiation  being 
dominant emission mechanisms. This is an extension of the preliminary results
of Dyks, Rudak \& Bulik \shortcite{bulik}.
In section 2 we outline the model and introduce 
the input parameters for which Monte Carlo simulations were
performed. Section~3
describes the numerical results and offers their interpretation; 
conclusions follow in section 4.

\section{The Model}

The presence of two peaks with large (0.4 - 0.5) 
phase separation in gamma-ray lightcurves within single polar cap models 
requires a nearly aligned rotator
(e.g. Daugherty \& Harding 1994) 
where the following three characteristic 
angles are to be of the same order: $\alpha$ - 
the angle  between the spin axis $\bld \Omega$ 
and the magnetic moment $\bld \mu$, 
$\vartheta_\gamma$ - the opening angle between the direction of 
the gamma-ray emission and $\bld \mu$,
and $\zeta$ - the angle between $\bld \Omega$ and the line of sight. 
For a canonical polar cap and instant electron acceleration, $\vartheta_\gamma$
roughly equals $0.02/\sqrt{P}$ radians only (where $P$ denotes a spin period).
To avoid uncomfortably small characteristic angles,
Daugherty \& Harding \shortcite{dh96}, postulated
that primary electrons come from extended polar caps,
and with the acceleration occuring at a height $h$ of several neutron-star 
radii $R_{\rm ns}$.
The latter assumption may be supported by the results 
of Harding \& Muslimov \shortcite{ham}
who investigated in a self-consistent way particle acceleration by the electrostatic field
due to field-line curvature \cite{arons}
and inertial frame dragging effect 
\cite{muslimov}. Harding \& Muslimov found that a stable
accelerator, with double pair-formation-front
controlled by curvature radiation,
is possible at a height of about 0.5 to 1 stellar radii.
Here we use a polar cap model combined with the assumption of a nearly aligned rotator.
Most ingredients of the pc-model come from
Daugherty \& Harding \shortcite{dh82}. Geometry of the magnetic field of a neutron
star is assumed to be well described by a static, axisymmetric dipole.

\begin{table*}
\begin{minipage}{\hsize}
\caption{Model parameters.}
\label{tab1}
\begin{tabular}{c c c c c c}
\hline 
    & \above{B_{\rm pc}}{[10^{12} \rm G]} & \above{\alpha}{[\rm deg]} 
       & \above{h_{\rm init}}{[R_{\rm ns}]} 
       & \above{\theta_{\rm init}}{[\theta_{\rm pc}]} & primary electrons \\

Model A & 1.0 & 3.0 & 0.0 & 1.0 & $E_{\rm init}=8.68$ TeV, no acceleration\\
Model B & 1.0 & 5.0 & 1.0 & 1.0 & $E_{\rm init}=20.0$ TeV, no acceleration\\
Model C & 3.0 & 10.0 & 2.0 & 2.0 &
$E_{\rm init}=0.5\MeV$, acceleration (see eq.(\ref{eq1}))\\ 
& & & & & with $V_0 = 1.5\times 10^{13}$Volts\\
Model D & 1.0 & 3.0 & 0.0 & $[0,1]$ & 
$E_{\rm init}=8.68$\ \rm TeV,\ no\ acceleration\\
\hline
\end{tabular}
\end{minipage}
\end{table*}

\begin{figure}
\epsfig{file=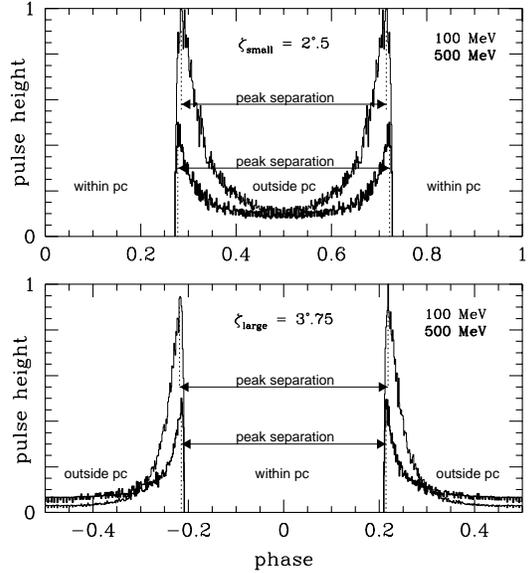, width=8cm}
\caption{
The double-peak pulses at 100 MeV (thin line)
and 500 MeV (thick line) calculated for model A. Depending on
viewing angle $\zeta$ between the spin axis $\bld \Omega$ and the line of sight
each single polar cap model yields two types
of double-peak pulses.
The upper and lower panels show the pulses for two values of the 
angle $\zeta$: $\zeta_{\rm small} = 2.5\,$deg and 
$\zeta_{\rm large} = 3.75\,$deg, respectively. 
}
\label{profiles}
\end{figure}

Within a given polar cap (pc) model with a fixed value of \al~ there are two possible
values of viewing angle $\zeta$ resulting in an identical peak 
separation $\del$  (defined as a fraction of $2\pi$) but with a reversed order of the two peaks 
(in terms of a leading peak, and then a trailing peak). 
Fig.\t \ref{profiles} illustrates the ambiguity in the definition of a double-peak pulse.
A simple geometrical relation
connects the three angles of interest - \al, \ze~ and $\thg$ - to the  phase separation 
$\del$:
\begin{equation}
\cos\thg = \cos\delta\sin\alpha\sin\zeta + \cos\alpha\cos\zeta,
\label{thg1}
\end{equation}
(eg. Ochelkov \& Usov 1980), 
where  $\delta = \pi\del$ for \ze$_{\rm large}$, and
$\delta = \pi(1 - \del)$ for \ze$_{\rm small}$
This relation holds as far as aberration of photon propagation due to rapid rotation is neglected.

Throughout the paper we always take the case of the larger $\zeta$ in each
model (4 models are considered). Its value ($\zeta_{\rm large} = 3.75, 4.5, 15.$, and $3.65$ 
degrees for models A, B, C, and D, respectively) along with our
choice for the angle \al~ (see Table 1) is to yield the separation $\del = 0.43$
at $300\MeV$.

The dominant HE emission process is
the curvature radiation (CR) by   
ultrarelativistic beam particles (primary electrons which leave the polar cap), 
followed by magnetic pair
production with the subsequent synchrotron emission (SR).
Our numerical code to follow the cascades induced by beam particles
is based on Daugherty \& Harding \shortcite{dh82}, and takes
advantage of the following approximations relevant to the
problem of directional and spectral distributions of the  photons:
The curvature photons are emitted tangentially 
to the local magnetic field direction in a frame rotating with the star.
The created e$^\pm$-pairs are assumed to follow the directions of their parent
CR photons and they share the photons' energy equally (for justification
see Daugherty \& Harding 1983). 
The synchrotron photons are emitted 
perpendicularly to the local magnetic field direction
in a frame comoving with electron/positron center of gyration. The pairs are assumed not
to change their position on the field line when radiating (for
magnetic field strengths considered here, energy-loss length scales due to SR
are of the order of $10^{-6}$ cm).
The emergent high-energy spectrum is a superposition of CR and SR. 
We follow Rudak \& Dyks \shortcite{rd99} to calculate detailed broad band energy
spectra of the high-energy emission. 

Beam particles are injected along magnetic field lines into
the magnetosphere either from the outer rim of the polar cap (hollow cone column)
or from the entire polar cap surface (filled column).
We use two simple models for their acceleration to ultrarelativistic energies: 
1) instant acceleration and
2) acceleration due to a uniform longitudinal electric field $\cal E$
over a scale height $\Delta h$.  
The pulse shapes as a function of
photon energy were calculated numerically for four sets of initial
parameters (hereafter called models A, B, C and D).
Table 1 features most important model parameters. 
In models A and B the primary electrons are distributed evenly along a
hollow cone formed by the magnetic field lines from 
the outer rim of a canonical polar cap,
i.e.\t with a magnetic colatitude $\theta_{\rm init} = \theta_{\rm pc}$, where
$\theta_{\rm pc} \simeq (2\pi \, \rns/
c\,P)^{1/2}$ radians at
the stellar surface level ($h = 0$). 
The beam particles are injected at
a height $h_{\rm init}$ with some initial
ultrarelativistic energy $E_{\rm init}$ 
(listed
in Table 1) and no subsequent acceleration takes place.
The main difference between models A and B is due to different values 
of $h_{\rm init}$ (equal to 0 and 1 $\rns$ respectively) which
result in different locations of origin of secondary particles.
Changing these locations is an easy way to modify spectral
properties of emergent radiation and enables to change (preferrably - 
to increase) the angle $\alpha$
as constrained by the observed energy-averaged peak
separation $\Delta^{\rm peak} \approx 0.43$ \cite{kanbach}.
In model C we assume
$\theta_{\rm init} = 2 \theta_{\rm pc}$. 
The primaries are injected at $h_{\rm init} = 2 \rns$ with 
$E_{\rm init} = mc^2$ and undergo acceleration
by a longitudinal electric field ${\cal E}$ present over 
a characteristic scale height $\Delta h = 0.6\, R_{\rm ns}$, 
resulting in a total potential drop $V_0$:
\begin{equation}
{\cal E} (h) = \cases{ V_0/\Delta h,  &for  $h_{\rm init} \leq h \leq (h_{\rm init}+\Delta h)$\cr
0,  &elsewhere.\cr }
\label{eq1}
\end{equation}
For comparison, we considered model D with 
a uniform electron distribution over 
the entire polar cap surface (i.e. $\theta_{\rm init} 
\in [\, 0, \theta_{\rm pc}]$).
All its remaining features are indentical to model A.

The values of $E_{\rm init}$ in models A, B and D, and the potential drop
$V_0$ in model C were chosen to yield similar number of secondary pairs 
--- about $10^3$ per beam particle. In all cases the spin period 
of the Vela pulsar $P = 0.089\,$s was assumed.

\section{Results}

\begin{figure*}
\epsfig{file=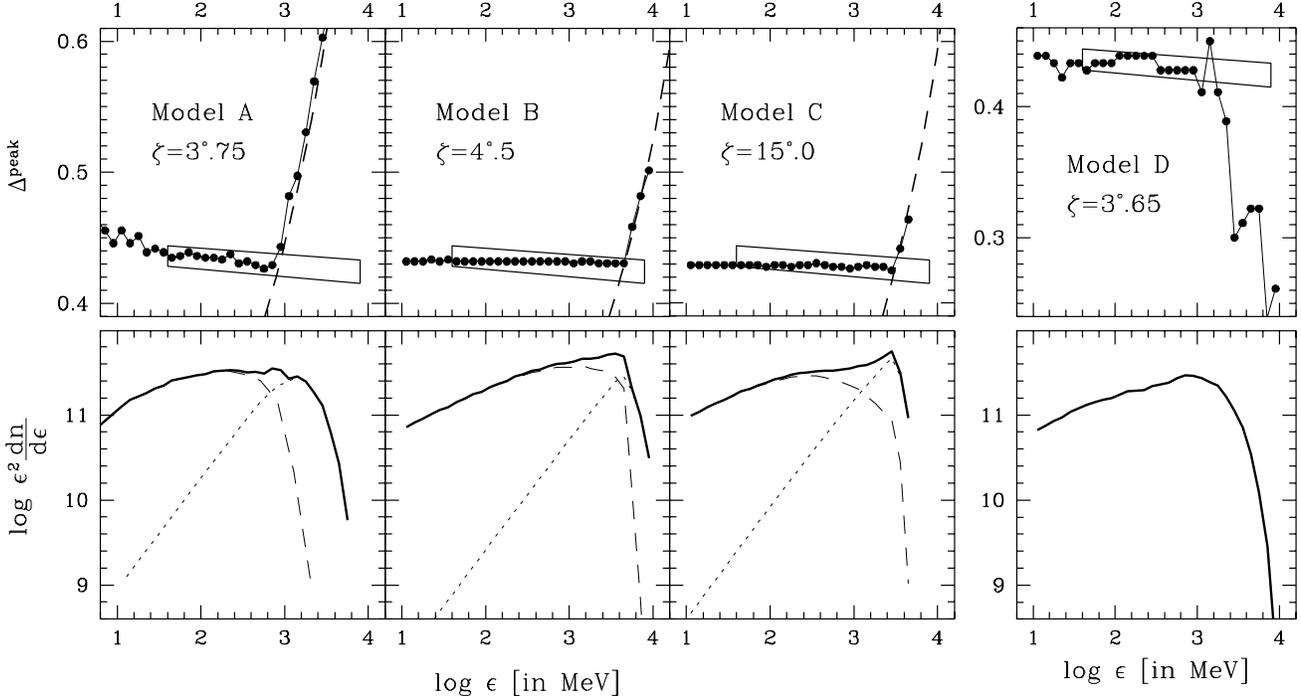, width=\hsize}
\caption{Upper panels:
Phase separation $\del$ versus photon energy $\varepsilon$ of the emission 
peaks found 
with Monte Carlo calculations for 
models A, B, C and D (dots). 
The overall observational trend, 
taking into account a substantial scatter of points (see fig.2, 
middle 
panel, of Kanbach 1999), 
is marked schematically for reference in all four panels
as a parallelogram. The assumed values of $\zeta$ - the angle between 
$\bld \Omega$ and the 
line of sight - are also indicated.
The steep dashed lines in the three left-hand side panels (hollow-column models) 
are the approximate analytical
solutions for $\del$ determined by magnetic absorption effects 
($\gamma \bld B \rightarrow e^\pm$). 
They reproduce the Monte Carlo results with good accuracy
in the high energy limit (the GeV domain).\hfill\break
Lower panels:
Energy output per logarithmic energy bandwidth 
at the first peak as a function of photon energy $\varepsilon$
for models A, B, C and D. The synchrotron (long dashed) and the curvature
(short dashed)
components are marked for models A, B, and C.
[Note: these are not instantaneous spectra.]
Units on the
vertical axis are arbitrary.
}
\label{sepspec}
\end{figure*}

General properties of the function $\del(\varepsilon)$ may be described in short in the following way:
In the low-energy range (i.e. below a few GeV)  the peak separation either 
remains constant with the increasing photon energy (Models B, C, and D) or
slightly decreases  (Model A). In either case, however,
the slope of $\del$ versus $\varepsilon$ looks
consistent with the results of Kanbach \shortcite{kanbach}.
Then, around a few GeV there exists a critical energy $\et$, at which 
the separation $\del$ undergoes a sudden turn: in the high-energy domain 
(i.e. for $\varepsilon > \varepsilon_{\rm turn}$) it increases in our hollow-column models (A, B, and C),
whereas it
decreases in the filled-column model (D), at a rate $\sim 0.28$ phase per decade
of photon energy.
The existence of $\et$  is a direct consequence of magnetic absorption 
($\gamma \bld B \rightarrow e^\pm$) in the magnetosphere (see also 
Miyazaki \& Takahara 1997). 
[Note, that $\et$ is not equivalent
to a high-energy cut-off  in a spectrum due to the magnetic absorption]. \hfill\break
The dependence of peak separation upon photon energy for all four models 
is presented in Fig.\t \ref{sepspec} (upper panels).

To understand why the slope in the low-energy domain ($\varepsilon < \varepsilon_{\rm turn}$) is different
in different models, we will now present several factors which  
are of interest in this respect.

Let us start with possible consequences of the orientation on $\del(\varepsilon)$ in this context, i.e. with
the choice of the angles \al~ and \ze.
For 
a given emission pattern in a frame of magnetic dipole rotating with the
star and for a set of values
of angles $\alpha$ and $\zeta$ fulfilling the condition 
$\del(300\MeV)=0.43$, a line of view can cross
the hollow cone of emission at different impact angles. 
Let us
estimate how geometry alone would affect the relations 
$\del(\varepsilon)$ shown in Fig.~\ref{sepspec}.
We neglect aberration of photon propagation due to rotation when calculating
directional emission
pattern as seen from some inertial observer frame.
For $\del$ staying close to 0.5 (roughly between $\sim 0.3$ and $\sim 0.7$)
$\cos(\pi\del) \simeq \frac{\pi}{2} - \pi\del$, the value of $\cos\thg$ is
approximately proportional to $\del$ with a constant of proportionality equal
to $-\pi\sin\alpha\sin\zeta$. Moreover, for a radiating particle sliding along a dipolar field line 
with the dipole constant $k = r\,\sin^{-2}\theta$
($r$, $\theta$ are the coordinates of the particles in the dipolar frame), 
one can link the opening
angle $\thg$ with a radial position $r$ of the particle:

\begin{equation}
\cos\thg = \frac{2 - 3\frac{r}{k}}{\sqrt{4 - 3\frac{r}{k}}}.
\label{thg2}
\end{equation}
Since in polar cap models $\frac{r}{k} \ll 1$, the right-hand side of eq.\t (\ref{thg2}) 
can be approximated by $\cos\thg \simeq 1 -
\frac{9}{8}\frac{r}{k}$ which along with eq.(\ref{thg1}) gives

\begin{equation}
r \simeq \frac{8}{9} \pi a \del + b
\label{geom}
\end{equation}
where
\begin{displaymath}
a \equiv k \sin\alpha\sin\zeta,
\end{displaymath}
and
\begin{displaymath}
b \equiv \frac{8}{9} \, k\, (1-\frac{\pi}{2}\sin\alpha\sin\zeta -
\cos\alpha\cos\zeta).
\end{displaymath}
The slope of the relation $\del(\varepsilon)$ is a combination 
of two factors: 1) viewing geometry and 2) directional and spectral changes in the radiation yielding the peaks
with increasing 
radial coordinate $r$:
\begin{equation}
{d \del \over d \varepsilon} = {d \del \over 
d r} \cdot {d r \over d \varepsilon}.
\label{slope}
\end{equation}
Any changes in viewing geometry
(i.e. the angles \al~ and \ze) will affect the slope $d \del / d \varepsilon$
via $d \del / d r = \frac{9}{8 \pi a }$, the latter being calculated from eq.(\ref{geom}).
For example, for models A and B the parameter $a$ equals $1.46 \rns$ and $2.91 \rns$, respectively.
Should $d r / d \varepsilon$ be identical for these two models, 
the slope in model A would be about two times bigger than in model B.
Fig.\t \ref{sepspec} (upper panels) does not show any such effects: the slope in model B
is $\approx 0$ for $\varepsilon < \varepsilon_{\rm turn}$.
This is because the viewing geometry effects are dwarfed
by  differences in directional and spectral properties of the radiation
in these models.

\begin{figure*}
\epsfig{file=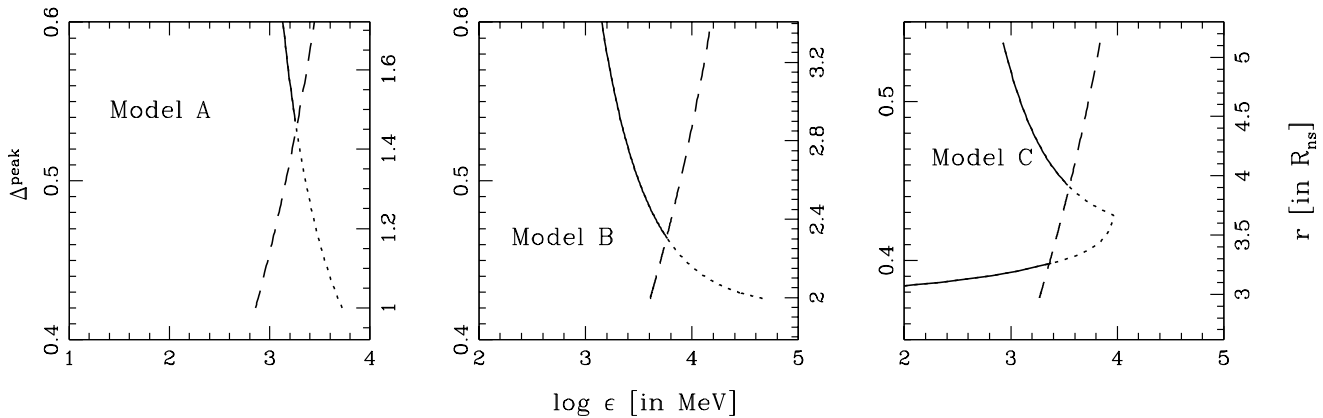, width=\hsize}
\caption{
Phase separation $\del$ versus photon energy $\varepsilon$ 
corresponding to the directions of propagation of curvature photons emitted by electrons
located on annuli (each annulus 
corresponds to a given radial coordinate and a dipolar constant; 
it translates
into $\del$ for a fixed set of angles between 
the spin axis $\bld \Omega$, 
the  magnetic moment $\bld \mu$, and the line of sight).
The curvature photons  are here assumed to have a characteristic energy $\ecr = 1.5 c \hbar\gamma_{\rm el}^3 
\rho_{\rm cr}^{-1}$ (the dimensionless energy $\gamma_{\rm el}$ of the electrons 
as well as the local radius of curvature $\rho_{\rm cr}$ change along electron's trajectory 
in a model dependent way). 
Dashed lines (as in Fig.~\ref{sepspec}, upper panels) represent the limiting energy $\ee$ of curvature photons
for which the magnetospere
becomes opaque. Photons located to the right of the dashed line in each panel
convert into $e^\pm$ pairs.
This part of $\del$ vs. $\varepsilon$ is shown as dotted line.
The pairs produce the synchrotron spectral component which in turn determines
the character of $\del$ vs. $\varepsilon$ in the low-energy domain 
(see Fig.~\ref{sepspec}, upper panels).
Note a wide range of $\del$ values for which curvature photons are absorbed 
in Model A, in contrast to Models B and C. 
}
\label{f2simp}
\end{figure*}

To analyse directional and spectral properties of the peak emission it is instructive
to begin with a simplified, one-component model.
Suppose that the only contribution to the emission is due to optically thin CR, i.e. 
both magnetic absorption and synchrotron emission
by secondary pairs created due to this absorption
are neglected. Numerical simulations of this case (not included in this paper)
show that
the peak separation in such a case, which we denote as $\delcr$, stays constant as a function of $\varepsilon$
for the models with instant acceleration (A, B, D), while it
increases with increasing $\varepsilon$ for model C (acceleration
over $\Delta h$) at a rate dependent on 
the electron acceleration rate.
The former case can be understood by arguing
that energy $E =
\gamma mc^2$ of the electron  decreases monotonically from its starting value $E_{\rm init}$, while
its radius of curvature $\rho_{\rm cr}$ increases. Recalling the properties
of the curvature continuous spectrum due to a single particle 
(e.g. Landau \& Lifshitz 1973) 
this makes the contribution to the spectrum per unit distance by the electron
to be the highest one just at the initial altitude $h_{\rm init}$.
In consequence, the contribution to the spectrum per unit phase angle by 
a bundle of electrons moving along a set of open field lines is the highest one at
$h_{\rm init}$.
Therefore, it is the opening angle of the CR photons at $h_{\rm init}$ which fixes the phase separation
$\delcr$ of the two peaks regardless the photon energy. 
In the latter case
the qualitative behaviour of $\Delta^{\rm peak}_{\rm
cr}(\varepsilon)$ can be understood by assuming that 
the phase of the pulse peak
at a given photon energy $\varepsilon$ is approximately 
determined by such altitude at which accelerated electrons reach
the energy $\gamma mc^2$ which satisfies the condition
$0.29\,\ecr(\gamma, \rho_{\rm cr}) \approx \varepsilon$, where
$\ecr = {3 \over 2} c \hbar\gamma^3 
\rho_{\rm cr}^{-1}$ is a characteristic
energy of the CR spectrum and $\rho_{\rm cr}$ is a local radius of curvature; both $\gamma$
and $\rho_{\rm cr}$ are functions of altitude $h$.

We now relax these two simplifications and present the consequences. 
First, the inclusion of magnetic absorption
results in a strong response of $\delcr$ at highest photon energies - between some
critical energy $\et$ and the high-energy cutoff, as mentioned in the first paragraph. 
This effect will be addressed in the final part of
this section.
Second, adding up synchrotron component (SR) due to $e^\pm$ pairs changes the properties
of high-energy emission for $\varepsilon < \et$ (i.e. below a few GeV) significantly. 
Most notable is the domination of SR over CR in terms of intensity.
The total energy output per logarithmic energy bandwidth at the first peak as a
function of photon energy $\varepsilon$ for models A, B, C and D is presented
in Fig.~\ref{sepspec} (lower panels). Both components -- SR and CR -- are marked
schematically to show their relative importance.
Consequently, $\del$ is affected by 
directional and spectral properties of the SR emission.

The behaviour of $\Delta^{\rm peak}$ is shown in Fig.~\ref{sepspec} (upper panels).
Its slight decrease
with increasing $\varepsilon$ (model A) or no change at all (models B, C, and
D) - is due to a combination of factors which determine
the directional and spectral properties of SR.    
These include
energy and pitch angle distributions of secondary $e^\pm$ pairs,
as well as their vertical spread within the magnetosphere combined with a strength of the local magnetic 
field.
These factors change from one model to another:

\noindent
Model A - In strong local magnetic fields ($B_{\rm local} \approx 10^{12}$ G), 
efficient pair production requires lower electron energies
than in a low-field case (like in model B). The production
occurs over a wider range of altitudes.
Consequently, the spectrum of SR contributed locally by the pairs
evolves considerably over this range of altitudes. 
Lorentz factors of gyration $\gper$ are of low values (of the order of
$mc^2 /(15\eb)$, where $\eb \equiv mc^2 B_{\rm local}/\bcr$). The resulting 
local SR spectra are, therefore, very narrow 
(see also the lower panel of fig.1 in 
Rudak \& Dyks 1999): 
they spread between the (local) characteristic SR energy $\esr$ and  
the (local) cyclotron
turnover energy $\ect= 1.5 \eb / \sin{\psi}$ (where $\psi$ is the pitch
angle) which roughly  satisfy $\esr / \ect \approx \gper^2$, and the ratio does not exceed $10^3$ in model A.
With increasing height (therefore, with decreasing $B_{\rm local}$)
both $\esr$ and $\ect$ move towards lower and lower values. The final effect of this
softening is then a notable decrease of
$\del$ with increasing photon energy (Fig.~\ref{sepspec}, upper panel, Model A). 
In even stronger local magnetic fields (but not exceeding $\bcr$)
the rate of decrease of $\del$ with increasing $\varepsilon$ may be 
much larger, because $\epm$-pairs are produced with extremely low $\gper$
and synchrotron/cyclotron photons emitted at any altitude 
concentrate near the local cyclotron energy $\eb$.

\noindent
Model B - When cascades are to develop in a relatively weak magnetic field,
$B_{\rm local} \simless 10^{11}$ G, very high electron energy $E_{\rm init}$ is required.
The high value of $E_{\rm init}$ means rapid CR cooling which brings the electron energy
quickly below the level required for pair production. Curvature photons become too soft
for pair
creation via magnetic absorption very quickly after injection.
The bulk of pairs is created by electrons with their energies confined to
a narrow range at $E_{\rm init}$ and consequently the SR component originates within a narrow
range of magnetospheric altitudes (radial positions  $r$). 
The resulting $\del$ does not change with photon energy $\varepsilon$.

The difference between model A and B in radial extension $r$
of the regions of origin of curvature photons which are absorbed, producing pairs and eventually 
SR, is presented in Fig.\t \ref{f2simp} (right-hand vertical axes). The difference
is more appealing when presented in terms  of $\del$ (left-hand vertical axes).

\noindent
Model C - After initial stage of linear acceleration, the electrons enter a regime of
`radiation reaction limited acceleration'.
Over a considerable range of altitudes ($3.3 \rns \simless r \simless 3.6 \rns$)
the electrons' energy remains approximately constant and so does the pair production
efficiency (it is equal to $\sim 4\cdot10^{-3}$ pairs per
centimetre of the primary electron path). 
In such conditions, any evolution of the SR-spectrum over this range might affect $\del$ as a function
of $\varepsilon$.
Nevertheless, no significant
changes occur in $\del(\varepsilon)$. This is because a single-particle
SR-spectrum almost does not change its shape within the range of altitudes 
with stable, strong pair production.
The balance between the acceleration rate and the CR-cooling rate
stabilizes $\ecr$ as well as $\esr$. Any spectral changes 
of SR in its low-energy part, nearby $\ect$, are not relevant in the context 
of gamma-rays, since they occur near 100 keV, 
i.e. well below the energy range of \egret.
\footnote{The behaviour of SR in its low-energy part (i.e. in hard X-rays)
may  then lead to increase of $\del$ in the hard X-ray domain in comparison to $\del$ 
in the gamma-rays.
Assuming alternative  to our choice definition of $\del$, with $\zeta_{\rm small}$ (see Section~2),
the behaviour of $\del$ would be a mirror-reflection of our case. Then
this effect might explain low value of $\del$ within
the $2 - 30$ keV range found by Strickman, Harding \& deJager
\shortcite{strickman} in {\it RXTE\ } data of Vela.
Moreover, below  2~keV the synchrotron
component in the model spectrum drops below the level of CR-emission, 
and the value of $\del$ as for the gamma-rays is expected to be resumed there.}

\noindent
Model D (a uniform distribution of primary electrons over the polar cap but otherwise 
identical
to model A)  - the final effect does not resemble the decrease of
$\del$ with increasing photon energy found for model A. Apart from numerical fluctuations (see 
Fig.~\ref{sepspec}, upper panel) 
the separation $\del$ remains approximately constant as a function of $\varepsilon$.

Finally, let us discuss the inclusion of magnetospheric opacity due to 
$\gamma \bld B \rightarrow e^\pm$.
This effect becomes important above $\sim 1\GeV$, where most power is due to CR in all models.
In this regime the position of each peak in the pulse
is determined by magnetic absorption, and this 
results in a strong response of $\del$  between
$\et \simgreat 1\GeV$ and a high-energy cutoff:
At $\varepsilon = \et$ the separation $\del$ 
undergoes a sudden turn and starts
increasing (or decreasing) rapidly for $\varepsilon > \et$. 
For the hollow-column models (A, B and C)
the photons in both peaks of a pulse come from low magnetospheric altitudes 
with narrow opening angles. When 
$\varepsilon$ is high enough these photons will be absorbed.
Photons which now found themselves in the `new' peaks come from
higher altitudes (the magnetosphere is transparent to them) and 
have wider opening angles.  
In other words, inner parts of the `original' peaks in the pulse will be 
eaten-up and the gap between the peaks,
i.e. the peak separation $\del$, will increase (Fig.\ref{sepspec}). 
Our Monte Carlo results for $\del$ at $\varepsilon > \et$ may be reproduced with 
good accuracy by a simple analytical solution of $\tau_{\gamma B} \geq 1$ which has to be 
combined with eq.(\ref{geom}).
The requirement $\tau_{\gamma B} \geq 1$ is particularly simple
(with some well known approximations being used) since it refers to a photon created with a momentum
tangential to local dipolar magnetic field line at a height $h$ above
the neutron star surface (at radial coordinate $r = h + \rns$). The photon will undergo magnetic absorption 
(to be more precise, with a probability of $[1 - \exp (- \tau_{\gamma B})]$) if its
energy $\varepsilon$ satisfies the following condition:
\begin{equation}
\varepsilon > \ee = 7.6 \cdot 10^2 \left(\frac{P}{0.1\rm s}\right)^{1/2} \left(\frac{\bpc}{10^{12}\ \rm G}\right)^{-1}
\left(\frac{r}{\rns}\right)^{5/2} \ \ \rm MeV,
\label{esce}
\end{equation}
(cf. eq.\t 11 in Bulik et al. 1999). 
This formula is valid for hollow-column models but may be used for any dipolar field line (the factor $P^{1/2}$
comes just from choosing  the outer rim of 
a polar cap to be the site of field-line footpoints).
Although the pulsar spin (which leads to an 
aberration and a slippage of magnetic field under the photon's path)
has been neglected in derivation of eq.\t (\ref{esce}), the
formula gives $\ee$ in excellent agreement with Monte Carlo method for
the emission regions placed up to several $\rns$ above the surface and
rotation periods typical for strong-field pulsars ($\ga 10^{-2}$ s).
One may look at the eq.(\ref{esce}) as the condition for a radius of escape $r_{\rm esc}$
at a given energy $\varepsilon$. 
Photons of energy $\varepsilon$ will escape the magnetosphere only if they are 
emitted at $r\ge r_{\rm esc}$
which satisfies
$\varepsilon_{\rm esc}(r) \geq \varepsilon$.
The radial coordinate $r_{\rm esc}(\varepsilon)$ has to be combined now with eq.(\ref{geom}) to
give $\del$ relevant for the `magnetic absorption 'regime. This analytical $\del$ is shown
as dashed lines in Fig.\ref{sepspec} (upper panels) and in Fig.\ref{f2simp},
whereas the filled dots 
are the Monte Carlo results.
This branch of solution intersects the horizontal line
set by $\Delta^{\rm peak}=0.43$ at $\varepsilon_{\rm turn}\simeq \, 0.9$, $4.5$, 
and $3$ GeV for models A, B, and C, respectively.

For model D, with a uniform distribution of primary electrons over the polar cap (but otherwise 
identical
to model A) the changes of $\del$ above $\et$, occur in the opposite
sense. Unlike in previous models, here  both peaks of the pulse
are formed by
photons emitted tangentially to magnetic field lines attached to the polar cap
at some opening angle $\theta_{\rm init} < \theta_{\rm pc}$ . These photons 
are less attenuated than those coming from a hollow-column,
and in consequence the peak separation drops. A similar behaviour 
was obtained by Miyazaki \& Takahara \shortcite{miyazaki}, 
in their model of a homogeneous polar-cap.

We have also investigated the behaviour od $\del(\varepsilon)$ above $\et$ for
other distributions of primary electrons over the polar cap. We have
considered intermediate cases between models A and D, (i.e.\t with 
a uniform coverage of only an outer part of the polar cap area
between some inner radius $r_{\rm in} < \rpc$ and the polar cap radius $\rpc$),
 and models with
uniformly filled interior of the polar cap surface but with increased electron
density along the polar cap rim (cf.\t Daugherty \& Harding 1996).
We conclude that
regardless the actual shape of the active part  (i.e.\t `covered' with primary electrons)
of the polar cap (either an outer rim,
or an entire cap, or a ring, or entire cap/ring + increased rim density),
one does expect in general strong changes in the peak separation 
to occur at photon energies close to high-energy spectral cutoff due to 
magnetic absorption.

A word of technical comment seems to be appropriate for
$\del = 0.43$
at $\sim 300\MeV$.
It appears that a technique adopted by Kanbach \shortcite{kanbach} of fitting the observed
pulse shapes  with asymetric Lorentz profiles tends to overestimate the true
value of $\del$ by a few thousandth parts of phase. Therefore the actual value
may be closer to 0.42 than 0.43 (Maurice Gros, private communication).
Nonetheless, this shift in $\del$ does not change any conclusions of this
work.  

\section{Summary}

In this paper we addressed a recent suggestion of Kanbach
\shortcite{kanbach} that peak separation $\del$
in the double-peak gamma-ray pulses of the Vela pulsar may monotonically decrease
with increasing photon energy  at a rate $\sim 0.025$ phase per decade in energy
over the range $50\MeV$ to $9\GeV$, 
We calculated 
gamma-ray pulses expected in 
polar-cap models
with magnetospheric activity induced by curvature radiation of
beam particles.   Two types of geometry of 
magnetospheric column above the polar cap were assumed: a hollow-column 
associated with an outer rim of the polar cap
and a filled column associated with a uniform polar cap. Four models were considered with two scenarios for
the acceleration of beam particles.
Pulsed emission in the models was a superposition
of curvature radiation due to beam particles
and synchrotron radiation due to secondary $e^\pm$ pairs in magnetospheric cascades. 
The changes in the peak separation were investigated with Monte Carlo numerical simulations.

We found that regardless the differences in the models, 
the peak separation $\del$ below a few GeV,
where the emission is dominated by synchrotron component,
is either a weak decreasing function of photon energy $\varepsilon$,  or remains
constant. 
Both variants may be considered to be in agreement with the results of 
Kanbach \shortcite{kanbach}
for the latter are affected by large statistical errors.
A particular behaviour of $\del$ depends on a combination of
several factors,
including strength of magnetic field in the region of pair formation and 
model of electron acceleration (both of which determine spectral and directional properties 
of the radiation at different altitudes),
as well as viewing geometry.
Essentially, in strong fields, $B_{\rm local} \ga 10^{12}\G$, 
$\del$ decreases
with increasing photon energy $\varepsilon$, whereas
for $B_{\rm local} < 10^{12}\G$, the peak separation $\del$
stays at a constant level.

Moreover, we found that due to the magnetic absorption 
($\gamma \bld B \rightarrow e^\pm$) there exists a critical energy
$\et$ at which the peak separation 
$\del$ makes an abrupt turn and then changes dramatically
for $\varepsilon > \et$.
It increases in the hollow-column models (A, B, and C)
and decreases in the filled-column model (D), 
at a rate $\sim 0.28$ phase per decade
of photon energy. 
The numerical behaviour of $\Delta^{\rm peak}$ in this regime in the hollow-column models 
was easily reproduced to high accuracy with a simple 
analytical model of magnetospheric transparency 
for a photon
of energy $\varepsilon$, and its momentum
tangential to local dipolar magnetic field line at a site of its origin.
An exact value of $\varepsilon_{\rm turn}$ 
is model-dependent but it is  confined to a range
between $\sim 0.9\GeV$ and $\sim 4.5\GeV$.

To find such a hypothetical turnover of $\Delta^{\rm peak}$
in real observational data
would require, however,
high-sensitivity detectors, since for $\varepsilon > \varepsilon_{\rm turn}$ the expected
flux of gamma-rays drops significantly.
If detected, this turnover would be an important signature of polar cap activity in gamma-ray
pulsars. It would support 
the notion
that
high-energy cutoffs in gamma-ray spectra of pulsars are due to magnetic absorption.

The CR-induced cascades models, like those considered in this work, are not the only
possibility for nearly aligned rotators to produce double-peak pulses with large phase separations.  
There exists an alternative class of models - with pair cascades above polar cap 
induced by magnetic inverse Compton scatterings (ICS) of primary electrons in the field of soft photons 
from the stellar surface - proposed in a series of papers (e.g. Sturner \& Dermer 1994,
Sturner et al. 1995). 
In particular, Sturner et al. (1995) present the detailed Monte Carlo model spectra of the Vela pulsar.
They also present pulse profiles at a fixed energy
of 50 MeV (for several viewing angles) but no word of comment is given regarding the 
problem of $\Delta^{\rm peak}$ versus photon energy.
We expect the outcome to be qualitatively similar to our results.
First, the scatterings take place mostly within a very limited height above the
polar cap surface (below $h \sim R_{\rm pc}$) and the preferred directions
of propagation of the ICS photons will be fixed by magnetic field lines just above the surface.  
Therefore, 
$\Delta^{\rm peak}$ due solely to ICS photons should stay constant for a wide range of energy.
Inclusion of synchrotron photons due to pairs is unlikely to notably affect $\Delta^{\rm peak}$
unless the pair formation front is vertically more extended than for CR-induced
cascades.
Second, some turnover point at
$\varepsilon_{\rm turn}$ not exceeding 1~GeV should be present due to 
magnetic absorption. 
The behaviour of $\del$ for $\varepsilon > \et$ should roughly follow the dashed lines 
in Fig. 2 (upper panel) and Fig.3 as long as the assumption about photons (which are 
to be absorbed) propagating tangentially to local dipolar magnetic field line at
their site of origin 
remains valid for majority of ICS photons.\hfill\break
To verify this qualitative picture  
would, however, require detailed numerical
calculations.

\section*{ACKNOWLEDGEMENTS}
This work has been financed by the KBN grants 2P03D-01016,
and 2P03D-02117.
Support from Multiprocessor Systems Group at Nicholas Copernicus
University's Computer Centre
in providing facilities for the time/memory-consuming Monte Carlo
calculations is appreciated.
We are grateful to Gottfried Kanbach and Maurice Gros for valuable discussions
on processing and analysis of high-energy data for the Vela pulsar.
We thank the anonymous
referee for bringing our attention to the paper by Sturner et al. (1995).


\bsp 

\label{lastpage}

\end{document}